# Evaluation of large language models for discovery of gene set function

Mengzhou Hu[1†], Sahar Alkhairy[2†], Ingoo Lee[1], Rudolf T. Pillich[1], Dylan Fong[1], Kevin Smith[3], Robin Bachelder[1], Trey Ideker[1,2*], and Dexter Pratt[1*]

[1]*Department of Medicine, University of California San Diego, La Jolla, California, USA*

[2]*Department of Computer Science and Engineering, University of California San Diego, La Jolla, California, USA*

[3]*Department of Physics, University of California San Diego, La Jolla, California, USA*

[†]These authors contributed equally

[*]To whom correspondence should be addressed:

tideker@health.ucsd.edu, depratt@health.ucsd.edu

**Abstract**

Gene set analysis is a mainstay of functional genomics, but it relies on curated databases of gene functions that are incomplete. Here we evaluate five Large Language Models (LLMs) for their ability to discover the common biological functions represented by a gene set, substantiated by supporting rationale, citations and a confidence assessment. Benchmarking against canonical gene sets from the Gene Ontology, GPT-4 confidently recovered the curated name or a more general concept (73% of cases), while benchmarking against random gene sets correctly yielded zero confidence. Gemini-Pro and Mixtral-Instruct showed ability in naming but were falsely confident for random sets, whereas Llama2-70b had poor performance overall. In gene sets derived from 'omics data, GPT-4 identified novel functions not reported by classical functional enrichment (32% of cases), which independent review indicated were largely verifiable and not hallucinations. The ability to rapidly synthesize common gene functions positions LLMs as valuable 'omics assistants.

**Introduction**

A fundamental goal of the 'omics sciences is to identify the groups of genes responsible for all of the distinct biological functions of life, health and disease. In this vein, numerous mRNA expression experiments over the past several decades have produced sets of genes that are differentially expressed across conditions or that cluster by common expression patterns. Similarly, proteomics experiments produce clusters of proteins that are co-abundant, co-modified, or physically interacting; CRISPR gene knockout screens produce lists of genes required for fitness or a particular response; and so on. In all of these cases, the basic premise is that the identified genes work coherently towards the same biological process or function.

A common approach to interpret the genes identified in 'omics experiments is through functional enrichment analysis[1–9]. This method seeks to identify similarities between a cluster of 'omics genes and those from a large pre-defined collection of gene sets organized by shared functions or pathway categories[10–15]. This pre-defined collection can come from literature-curated gene function databases, such as the Gene Ontology[16,17], Kyoto Encyclopedia of Genes and Genomes[18–20], or Reactome[11,21,22]. Alternatively, one can perform enrichment analysis against databases of genes annotated from previous independent experiments, such as genes previously linked to the same disease in the Genome-Wide Association Studies Catalog[23], genes linked to the same mouse knockout phenotypes in the Mouse Genome Database[24,25], genes regulated by a common transcription factor[26,27], or genes that serve as canonical biomarkers for a given cell type[28–30].

Paradoxically, an 'omics gene cluster that is highly similar to one or more gene sets in a reference database may be of lesser interest, since the cluster and its function have already been well characterized. Of greater interest are clusters of genes not previously implicated together, because it is precisely in these cases that new biological insights emerge. These less-studied cases may either show no significant enrichment in the reference database, or they may return enrichments that are significant in terms of p-value, but not substantial in terms of overlap. Here, an immediate next step is to explore the biological literature, as well as complementary data sets, to learn as much as possible about the genes in question. The goal is to mine knowledge pertinent to each gene and then use this knowledge to synthesize mechanistic hypotheses for a function that might be held in common by all or most genes in the set. This protracted process of discerning relevant findings from data and literature, then reasoning on this information to synthesize functional hypotheses, has not yet been widely automated but is one of the central tasks performed by a genome scientist.

The advent of generative artificial intelligence (AI) models and, specifically, Large Language Models (LLMs) is highly relevant to these tasks. At its core, generative AI is an approach to machine learning by which a model is trained to recognize underlying patterns in data in a manner that allows it to generate new data with similar properties as the training data. The underlying technology behind LLMs is the transformer architecture[31–33], which uses a self-attention mechanism to understand context and handle long-range dependencies in text, delivering significant advancements in tasks such as text translation, summarization, and generation. Recent AI research has produced a flurry of general-purpose LLMs, such as GPT-4[34] by OpenAI, Llama2[35] by Meta, Mixtral[36] by Mistral AI, and Gemini[37] by Google, which incorporate information from an enormous corpus of sources, including the biomedical literature. Based on these developments, we hypothesized that LLMs might provide a major opportunity to assist in the interpretation of gene sets derived from 'omics experiments.

Here, we evaluate the degree to which LLMs provide insightful functional analyses of gene sets based on their embedded biological knowledge and text-generation capabilities. First, we develop a gene set analysis pipeline based on queries to a panel of current LLMs. We then test the ability of each LLM to propose succinct names describing the functions of gene sets of interest, as well as to support this choice by referenced text and an overall assessment of confidence. Finally, we discuss our findings and their implications for the general use of LLMs in functional genomics applications.

## Results

**Development of an LLM functional genomics pipeline.** We designed a pipeline in which an LLM is instructed to analyze a gene set and then generate a short biologically descriptive name, a supporting analysis essay, and a score reflecting the LLM's "confidence" in these results (**Fig. 1a, Methods**). A separate LLM instruction was used to validate statements made in the analysis essay with pertinent literature citations (**Extended Data Fig. 1, Methods**). The instruction to an LLM is called a "prompt" and can include data and examples to guide the response. Best practices for formulating this prompt are the subject of ongoing experimentation[38–41], but here our prompt simply described desired properties of the results to be generated, including guiding phrases such as "For each important point, describe your reasoning and supporting information." The engineered prompt also included a single (one-shot) example to help the LLM imitate the desired format and thought process (**Fig. 1a**, **Extended Data Fig. 1b, Extended Data Table 1**). Gene set query functionality with all tested LLMs is available for general use via the Gene Set AI (GSAI) web portal (https://gsai.ideker.ucsd.edu/).

We sought to evaluate this pipeline using reference gene sets derived from two primary sources. The first source was literature curation, for which we evaluated sets of genes drawn from Gene Ontology (GO) terms[16,17] (Evaluation Task 1; **Fig. 1b**). The second data source was 'omics analysis, for which we evaluated sets of genes identified by various 'omics platforms, including transcriptomics and proteomics (Evaluation Task 2, **Fig. 1c**). The goal of the first task was to benchmark how well LLMs recover gene set functions previously documented by a human-curated reference database, while the goal of the second task was to explore the extent to which LLMs provide novel insights beyond what is obtained from such databases.

**Evaluation Task 1: Recovery of Literature-Curated Functions.** For the first task, we randomly sampled a representative corpus of terms from the GO Biological Process branch (GO-BP 2023-11-15 release; **Extended Data Fig. 2, Methods**). The gene set annotated to each term was used to prompt five different LLMs (**Fig. 1b**), after which the names suggested by the LLMs were compared to the term names assigned by the GO curators. In each case, performance was measured by the semantic similarity of the LLM name to the GO name. Semantic similarity[42] is a quantitative score (range [0, 1]) that measures the closeness in meaning of two words or phrases, regardless of whether those phrases involve different words or expressions (**Methods**). For example, the word "socks" is semantically closer to the word "shoes" than it is to "airplane."

The five LLMs required from 7.9 seconds (Gemini Pro) to 61.8 seconds (Llama2 70b) to process a gene set and return a proposed concise name, a confidence score and supporting analysis text (**Extended Data Table 2**). Semantic similarity scores ranged from values as high as 1.0, in cases where the LLM name exactly matched the GO name (e.g. Gemini Pro: "Synaptic vesicle exocytosis", GO:0016079), to values below 0.1, in cases where the names were not intuitively similar (e.g., GPT-3.5: "Regulation of ion transport and cellular homeostasis" versus GO: "Negative regulation of CD8-positive, alpha-beta T cell differentiation", GO:0043377) (**Table 1**, **Supplementary Table 1**). We found that GPT-4, Gemini Pro, GPT-3.5 and Mixtral Instruct showed roughly equivalent performance in proposing a name that was similar to the GO name (median similarities in range 0.45 – 0.50), whereas the performance of Llama2 70b was significantly worse (median similarity = 0.40; **Fig. 2a**).

To interpret these similarity scores, we calibrated them against background semantic similarity distributions, defined by comparing each LLM-proposed name against the entire set of 11,943 term names documented in GO-BP (**Methods**). For example, the GPT-4 name ("DNA Damage Response and Repair") had a semantic similarity to the GO name ("Response to X-ray") of 0.54, a score that was higher than 99% of semantic similarities between the GPT-4 name and every other term name in GO-BP (**Fig. 2b, Supplementary Table 2**). Using this scoring approach, we found that 60% of gene set names proposed by GPT-4 were close matches to the corresponding GO term names, with semantic similarities ranking above the 95th percentile (**Figs. 2c, d**). In approximately one-third of remaining cases, the LLM proposed a name matching a more general ancestor in the GO term hierarchy (**Fig. 2d**). For example, the gene set corresponding to the GO term "Negative Regulation of Triglyceride Catabolic Process" resulted in the GPT-4 name "Lipid Metabolism and Trafficking" with a semantic similarity of 0.41 ranking in the 89th percentile. The GPT-4 name matched most closely to the GO term "Lipid Metabolic Process," a less specific category higher in the ontology and annotated by a larger set of genes (**Fig. 2e**). Qualitatively similar results were observed when analyzing gene sets from the Cellular Component and Molecular Function branches (**Extended Data Fig. 3, Supplemental Table 2**).

**Assessment of LLM confidence.** We next focused on the self-confidence reported by each LLM (**Extended Data Table 1**). As noted above (**Fig. 1a**), we had asked each LLM to provide a continuous confidence score[43,44] for each gene set analysis, in the range 0 to 1. For gene sets for which the LLM assigned a confidence of "0", we requested the LLM to return "System of unrelated proteins" rather than a proposed name, since it could not confidently propose a collective functional description. To gain insight into whether these confidence scores were informative and useful, we introduced in our evaluation the concept of contaminated gene sets. Specifically, each of 100 GO terms used previously was substituted for a synthetic gene set containing 50% of genes randomly selected from that GO term and 50% of genes randomly selected from the background pool of all genes with GO annotations ("50/50 mix", **Fig. 3a**). We also examined a fully random variant whereby 100% of genes were randomly selected from background ("Random"; **Fig. 3a**).

We observed that all LLM models, save for Llama2, showed a significant reduction in confidence when asked to generate names for the 50/50 mix and random gene sets (**Fig. 3b**). GPT-4 was the most likely of the five LLMs to correctly associate lower confidence with contaminated gene sets, and it gave zero confidence (refusing to name) the vast majority of gene sets that were fully random (87%). In contrast, GPT-4 assigned nearly all analyses involving real gene sets to a high or medium confidence (98%, **Fig. 3b**). We also compared these results to classic functional enrichment analysis run on the same real, contaminated, and random gene sets (BH-adjusted $p \leq 0.05$, **Methods**). As expected, enrichment analysis always returned the correct GO term for the real gene set, while for 85% of random gene sets, none of the enrichment results met the significance cutoff. However, enrichment analysis typically also returned significant GO terms for 50/50 mix (contaminated) gene sets (**Fig. 3b**). Thus, one might have expected the LLM to generate a name for any gene set, indiscriminately, but in this analysis it was more conservative in providing a name than classical functional enrichment.

**Evaluation Task 2: Exploration of gene sets discovered in 'omics data.** For the second task, we asked GPT-4 to name gene sets that had been identified experimentally, via clustering of 'omics data. These sets included: (1) Genes differentially expressed in transcriptomic profiles collected in response to a panel of drug treatments (n = 126, LINCS L1000 CMAP Signatures of Differentially Expressed Genes for Small Molecules) [45], (2) Genes differentially expressed upon infection by a panel of viruses (n = 48, GEO Signatures of Differentially Expressed Genes for Viral Infections) [46], and (3) Genes encoding complexes of interacting proteins identified by proteomic methods (n = 126, NeST) [47] (**Methods**, **Extended Data Fig. 4**). When prompted with each of these 300 'omics gene sets, we found that GPT-4 confidently annotated a name in 133 cases (44%, confidence ≥ 0.8) and otherwise deferred with zero confidence (**Table 2**).

To benchmark these results, we also subjected each gene set to classic functional enrichment analysis against the GO Biological Process database. Enriched GO terms were defined based on the significance of enrichment (**Methods,** BH-adjusted p ≤ 0.05) and the effect size, measured as the fraction of genes shared between the 'omics gene set and the GO term (Jaccard Index, JI). Notably, we found that even a very lenient overlap requirement (JI ≥ 10%) left 260 gene sets lacking annotation by GO terms (87%; **Table 2**, **Extended Data Fig. 5**, **Supplementary Table 3**). Of these non-enriched gene sets, 97 had been confidently processed by GPT-4 (37%, **Table 2**), yielding a novel functional name synthesized from outside of the GO corpus.

One example was NeST 2-123, a cluster of 13 biophysically interacting proteins identified through integrative proteomic mass spectrometry. Classical functional enrichment had returned no compelling hits, with the best matching GO term having only a single gene in common (JI = 0.06, BH-adjusted p = 0.04). Analysis of this cluster by GPT-4 generated the name "Endosomal Sorting and Trafficking" (**Table 3**), based on its finding that *"the majority of the proteins in this set are involved in the processes of endosomal sorting and trafficking… While not all proteins are directly linked to this process, the preponderance of sorting nexins and related proteins suggests a strong functional theme. The presence* *of proteins like REN and SNRPA1, which are not directly related to endosomal sorting, slightly lowers the confidence score. However, the overall function of the system appears to be centered on the endosomal-lysosomal pathway."*

**Assessment and validation of supporting analysis text.** Finally, we evaluated the analysis essays generated by GPT-4 in support of its proposed gene set names. We were specifically interested in the potential of LLMs to "hallucinate," i.e., to generate plausible but unverifiable or nonfactual statements[34], and the degree to which such hallucination might influence the gene set analyses here. For this purpose, four human scientists participated in a structured review process for 403 sentences generated in the analysis of 20 'omics gene sets (**Methods**). As a conservative criterion, we considered a sentence "verified" only if the reviewer found evidence in the literature for every stated fact. Of the 403 sentences evaluated, we found 354 to be fully verifiable (88%, **Supplementary Table 4**). Examination of the 49 remaining sentences revealed two major types of unverified facts: (1) Miscategorization of gene functions (n = 15, 4%) and (2) Speculation of gene functions (n = 34, 8%). In one case relevant to type (1), GPT-4 stated that WDTC1 *"is involved in the regulation of the cell cycle and apoptosis…"* when in fact, it is an E3 ubiquitin ligase and is involved in adipogenesis and obesity[48]. Relevant to type (2), the GPT-4 statement REN *"may be affected by vesicular trafficking processes"* could not be verified (paragraph 3, **Table 3**).

To facilitate statement verification, we developed a separate GPT-4-based system to add citations to the analysis essay in support of key statements made (**Extended Data Fig. 1**, **Supplementary Table 4**, **Methods**). In formulating the engineered prompt for this task, we did not stipulate that the title or abstract of a publication must be primarily about the statement; it was sufficient that a supporting fact was present. The 403 previously reviewed sentences returned 489 citations through this automated system. In 383/489 cases, the paper title or abstract provided clear evidence for the cited statement. For example, the statement that WDFY1 *"is implicated in signaling pathways, potentially acting as an adaptor in the endosomal system"* was supported in the titles of Hu et al. (2015)[49] and Ning et al. (2019)[50], as well as in the abstract of Yeo et al. (2019)[51] (see paragraph 9 and its citations in **Table 3**). The remaining 106 citations (22%) did not verifiably support their corresponding LLM statements, although we reviewed titles and abstracts only without systematic review of the manuscript main text. These results suggest that most but not all citations found by this procedure are reliable, such that they may be viewed as useful guidance for further study but not unquestioned facts.

## Discussion

The evaluations performed here suggest that LLMs have notable potential as automated assistants for understanding the collective functions of gene sets. In the analysis of gene sets from the Gene Ontology (GO), four out of five LLMs performed comparably in proposing names similar to the names assigned by the GO curators, with GPT-4 producing highly similar names for most gene sets. LLMs showed varying ability to score confidence in a proposed name, or to refuse to generate a name in cases of lowest confidence. The accompanying analysis text was found to be largely factual, although GPT-4's occasional generation of unverifiable statements shows that even current state-of-the-art LLMs should be coupled to fact-checking and/or reference validation, whether automated or manual.

When the GPT-4 name for a GO gene set was not similar to the curated name, in roughly a third of those cases it was conceptually broader (**Fig. 2d**). For the remaining gene sets with discrepant naming, it is possible the mismatch is due to a failure of GPT-4 to recover a well-documented common function or that the GO term no longer reflects the up-to-date literature. Alternatively, it is possible that both GPT-4 and GO offer valid, but alternate, interpretations. We indeed find anecdotal evidence for this last possibility: for example, Dendritic Cell Dendrite Assembly (GO:0097026) is annotated with two chemokines, CCL19 and CCL21, and their receptor CCR7, but these proteins are also critical to the related process of lymphocyte homing, consistent with the GPT-4 proposed name "Lymphocyte Homing and Immune Response Regulation." In other cases, the names are of different types, such as the high-level phenotypic GO term Reproductive Behavior (GO:0019098), which GPT-4 named Neurotransmission and Neuroendocrine Signaling, focusing instead on molecular and cellular processes (**Supplementary Table 2**).

The current state of the art, functional enrichment analysis, is a statistical method to quantify the agreement of a gene set with sets stored in fixed curated reference databases. In contrast, GPT-4 synthesizes common functions for genes based on active reasoning over a large corpus of biomedical knowledge, producing results even in cases where the gene set does not resemble any set previously recorded and ascribed a function. In particular, when analyzing experimentally derived 'omics gene sets, in about a third of cases we saw that GPT-4 was able to propose a name while functional enrichment found no matching term. An exciting possibility for future research would be to integrate the best of these two worlds, combining the precision of enrichment analysis with the literature knowledge and reasoning of LLMs.

Another axis for comparing the two approaches relates to transparency. On one hand, enrichment analysis could be said to be highly "transparent" because it uses well-defined statistical methods and documented reference databases of gene sets that researchers can review. In contrast, the knowledge accessed by the GPT-4 is not directly subject to definition or inspection, as it is embedded in the latent space of the model, and the mathematical calculation behind a given output is practically opaque. On the other hand, there is a sense in which an LLM analysis is extremely transparent because it presents a *de novo* narrative of facts and reasoning that considers all of the genes of interest, assisted by reference citation tools such as the one developed in this study. Ultimately, how an analysis was performed may be less relevant if the output is useful and its statements are supported by reasoned arguments and verifiable literature citations. Indeed, these are the criteria we apply to analyses produced by human researchers; we can see their outputs but not the operation of their minds.

Gene sets emerging from an experimental study do not always have a coherent function that can be summarized or serve as the basis of a name. Here we have seen that, via a self-confidence scoring measure, an LLM can assess the coherence of a gene set, potentially alerting biologists to cases in which they should be skeptical of a simple "best match" function proposal. For example, we found GPT-4 nearly always proposes a confident name for gene sets affiliated with terms in GO, while naming only about 13% of sets drawn from random genes and less confident in names for contaminated sets (**Fig. 3b**).

It is important to stress that our evaluation of the GPT-4 model was based on single queries using prompts developed by informal experimentation. Further research should investigate more sophisticated prompting strategies and methods that apply external tools and orchestrate multiple LLM interactions[52–58], such as integrating literature searches or gene set enrichment into the LLM analysis rather than as a post-hoc verification method. While incorporating techniques such as in-context learning, fine-tuning, and retrieval augmentation hold promise for enhancing the accuracy and interpretability of LLMs, these methods are not the focus of our current evaluation. Our explorations do, however, represent a future direction for research in genomic analysis and the development of AI models. Our study has also not attempted to augment LLM prompts with descriptions of the biological and experimental context in which a gene set was discovered, information that might improve the specificity, depth, and quality of the analysis. Future work could explore the inclusion of disease and experimental conditions in the prompt to enable the proposal of context-specific gene functions. Such prior context has been difficult to capture using gene set functional enrichment tools, since their pre-existing mapping of gene sets to functional terms is static and does not attempt to encode the practically infinite space of biological conditions.

In summary, one might have suspected that using LLMs to study gene function would produce statements, hypotheses, and references that hallucinate so uncontrollably as to be unusable. In fact, the more advanced models such as GPT-4 typically did not, showing reasonable and often exemplary performance over a series of complementary benchmarks. We thus conclude that, given appropriate framing, LLMs provide researchers with a new and powerful tool for gene set interpretation.

## Methods

**Large language model installation.** Five large language models were selected for the evaluation, including GPT-3.5 and GPT-4 from OpenAI, Gemini Pro from Google, Mixtral Instruct from MistralAI, and Llama2 70B from Meta. We used the 'gpt-4-1106-preview' and 'gpt-3.5-turbo-1106' versions of the OpenAI GPT-4 and GPT-3.5 large language models and the 'gemini-pro' version of the Google Gemini model using their well-defined APIs. Mixtral Instruct and Llama2 were downloaded from Ollama (ollama.com) and queried through the API endpoint of Ollama.

**Controlling the variability of LLM responses.** Each LLM enables queries to set a "temperature" parameter that controls the variability of the generated response, with lower temperatures producing more reproducible and reliable responses[59,60]. Exploring the effect of temperature on LLM analyses is outside the scope of this study, and therefore our queries used the lowest, most conservative/reproducible temperature value = 0.0. In a manual inspection of repeated queries at temperature 0.0, we found that LLM names and analyses were conceptually equivalent but that the specific text could vary, from near identity to considerable differences in phrasing. The 'seed' parameter was set to 42 for all models and all runs. Additionally, we made our manual review process manageable by forcing the responses to be concise. For this purpose, we set the maximum number of tokens (roughly corresponding to words) in each response to be 1,000.

**Prompt engineering.** The LLM prompt was organized in seven sections (**Fig. 1a,** see full prompt in **Extended Data Table 1**). System content section: System content tells the role of the LLM when to process the prompt. Here our analysis was associated with molecular biology, thus we set the role to be 'assistant of a molecular biologist'. Task instruction section: The instructions were engineered to meet multiple criteria. Notably, the LLMs were guided to first perform the analysis before proposing a process name, encouraging a structured "Chain-of-Thought." Confidence score assignment section: This prompt section instructed the LLM to generate a "confidence score" expressing its confidence in its choice of name, taking into account the fraction of genes that participate in the corresponding biological process(es). The coherence score was specified to be between 0.00 and 1.00. The prompt was also engineered to handle situations where the genes in a set are not sufficiently related to warrant a name. In particular, the prompt instructed the LLM to output a zero confidence score and the name "System of unrelated proteins" in these cases. <u>Format instruction section:</u> We asked the LLM to place the name as a title in the final analysis for easy extraction. <u>Analytical approach section:</u> The instructions in this section guided the LLM to be succinct, factual, and focused on finding commonalities and relationships. <u>One-shot example section:</u> This section contained an example of a gene set and the corresponding name, confidence score, and analysis text. This format follows the "in-context learning" approach, in which examples provide a template to help the LLM generate outputs consistent with the desired behavior and format. After substantial manual testing, we determined that the quality of the output was no different when using one example versus several examples; thus we chose to use a "one-shot" single example strategy, minimizing both prompt size and associated costs.

**Download and parse the Gene Ontology.** The Gene Ontology (2023-11-15 version) was obtained from the geneontology.org website in the Open Biomedical Ontologies (OBO)[61,62] format. The ontology file was subsequently divided into its three constituent branches: Biological Processes (BP), Cellular Component (CC), and Molecular Function (MF). The gene set corresponding to each GO term was determined by aggregating the genes with which it was directly annotated with those of all its ontological descendants.

**Calculation of semantic similarity.** Semantic similarity between names was determined using the SapBERT model[63]. SapBERT produces embeddings of each name and then computes the cosine similarity between the embeddings, yielding a similarity score ranging from 0 (no similarity) to 1 (identical). SapBERT is a domain-specific language representation model pre-trained on large-scale biomedical data, including Unified Medical Language System (UMLS), a massive collection of biomedical ontologies with 4M+ concepts. Since models like BERT are trained on vast amounts of textual data, they can learn general patterns and relationships and capture context by considering surrounding words, providing a measure of similarity based on semantics rather than lexical matching. Although both SapBERT and GPT-4 are LLMs, they are separate models with different purposes, model architecture, training objectives, and data. SapBERT therefore provides an independent evaluation of similarity.

**Calibrating the similarity between GPT-4 names and GO names.** To evaluate the performance of the GPT-4 model in recapitulating GO names, we computed the semantic similarity between the GPT-4 name and the assigned name of the GO term query, using SapBERT as described above. We then performed this semantic similarity calculation for the same GPT-4 name against every other GO term name in the Biological Process branch (GO-BP), yielding a background distribution of semantic similarity scores for each GO term query. The actual and background similarities were then concatenated into a single list, sorted in descending order (largest to smallest), and the rank of the actual similarity was recorded and expressed as a percentile. This percentile score is thus the percentage of GO-BP term names that are less similar to the GPT-4 name than to the assigned name of the GO term query.

**'Omics data processing.** For each 'omics source we selected gene sets with a size between 3 and 100 genes. Furthermore, in the L1000 dataset, we selected the context with the greatest number of observations {Cell line: "MCF7", duration: 6.0h, dosage: 10.0µm}. For the Viral disease perturbations dataset, we used a z-score cutoff of 2.

**Evaluation of GO enrichment.** We used the Enrichr web service to perform gene set enrichment in both Task 1 and Task 2. GO-BP terms for the queried gene set were obtained using the "gp.enrichr" function from GSEAPY package[64]. Task 1 utilized a custom input, the GO-BP (2023-11-15 version and Task 2 applied predefined parameters {gene_sets = 'GO_Biological_Process_2023', organism='human'}. In Task 2, we computed the Jaccard Index overlap between the gene set of interest and the GO term. For cases with multiple significant GO terms (adjusted p-value ≤ 0.05) we selected the term with the largest JI. A gene set was considered annotated by enrichment analysis if the adjusted p-value was ≤ 0.05 and the largest JI was ≥ 0.1.

**Evaluation of GPT-4 for 'omics gene sets.** For each 'omics gene set, we queried the GPT-4 pipeline (described above) for a name, analysis text, and confidence score. A gene set was considered annotated by GPT-4 if the confidence score was ≥ 0.8.

**Identification and validation of relevant references (citation module).** We followed a 5-step process to identify and evaluate references for statements made in the LLM-generated analysis text. For each paragraph in the analysis text, we performed the following (**Extended Data Fig. 1**):

1. Prompt the LLM to extract two types of keywords from the analysis paragraph: (1) gene symbols explicitly mentioned in the paragraph and (2) up to three keywords associated with gene functions or biological processes, ordered by their importance. Paragraphs that do not yield at least one gene symbol and one functional keyword are skipped, returning 'unknown.' The prompt incorporates a one-shot example of a paragraph and corresponding keywords.
2. Assemble a PubMed query expression to find scientific publications in which either the title or abstract contains one or more of the gene symbols and one or more of the function keywords.
3. Query PubMed via its web API, sorting the returned publication list by relevance.
4. Further prioritize the publications based on the number of matching genes in the abstract. We prefer publications that provide information on the most genes.
5. For each of the top three publications, prompt the LLM to assess whether the title and abstract provide evidence for one or more statements of fact in the analysis paragraph. Return the publication as a reference if the LLM considers that it satisfies that requirement.

**Reviewer fact-checking of GPT-4 analysis text.** We performed a structured review of 403 sentences from the analysis text generated by GPT-4 based on 20 selected 'omics gene sets (**Supplementary Table 3**). In this review, each of the four reviewers recorded the number of unverified statements of fact for each analysis in the corresponding column. A statement was considered "unverified" if no supporting evidence was found within roughly ten minutes, using the following method:

- Check simple per-gene statements against information from NCBI Gene content maintained by the National Library of Medicine, http://www.ncbi.nlm.nih.gov.
    a. For example, "Oxytocin (OXT) is a neuropeptide hormone that binds to its receptor, oxytocin receptor (OXTR)." can be quickly verified by the NCBI Gene entries for the two genes.

- 
    - 
        b. If the NCBI entry verifies one or more statements, add the URL for the entry to the Evidence column, e.g., "NLM: OXT http://www.ncbi.nlm.nih.gov/gene/5020"
    - For statements not verified by NCBI Gene, search PubMed for publications to provide evidence for the statement. Search strategies include:
        a. Search using gene-keyword pairs, such as "TP53 cell cycle".
        b. For paragraphs that discuss multiple genes, search for review articles with phrases such as "acute phase response proteins."
        c. Search for family member proteins together, such as "TAS2Rs bitter taste".

**Reviewer evaluation of references.** The reviewers evaluated references based on the same criteria with which the LLM was prompted in step 5 of the reference-finding process (above). Reviewers separately recorded whether the title or the abstract successfully provided evidence for a statement of fact, along with the number of irrelevant references for a paragraph.

# Figures and Tables

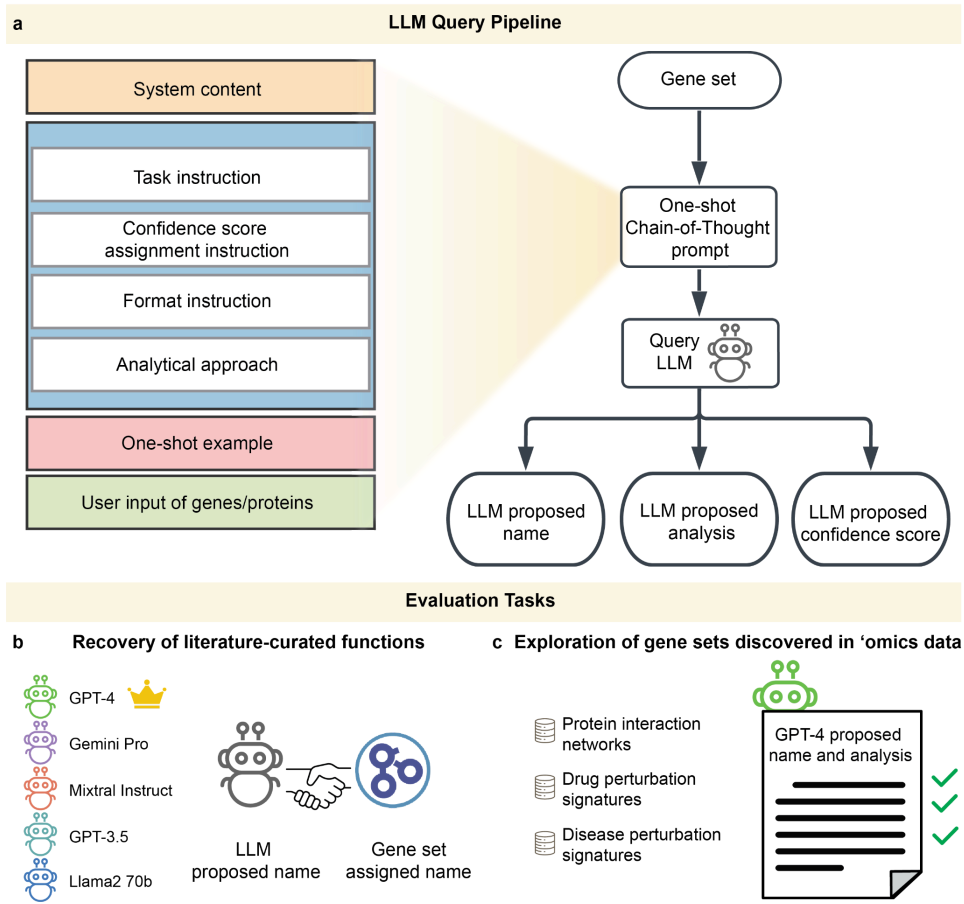

**Fig. 1: Use and evaluation of LLMs for functional analysis of gene sets. a**, The LLM prompt (left boxes) includes system content, detailed chain of thought instructions, and an example gene set query with desired response (full prompt given in **Extended Data Table 1**). The specific list of genes is inserted into the "User input of genes/proteins" field at the end of the prompt template, resulting in generation of a proposed name, a supporting analysis essay and a confidence score (right flowchart). **b,** Benchmarking LLM names against names assigned by GO (Evaluation Task 1). The proposed name from each of five LLMs (left robot icons) is compared to the name assigned by the GO curators (handshake icon). GPT-4 (crowned) was the winning model for this task. **c**, Exploration of gene sets discovered in 'omics data (Evaluation Task 2). The GPT-4 name and analysis are scored for novelty and accuracy (right green check marks). Gene sets derived from three different data types (left database icons).

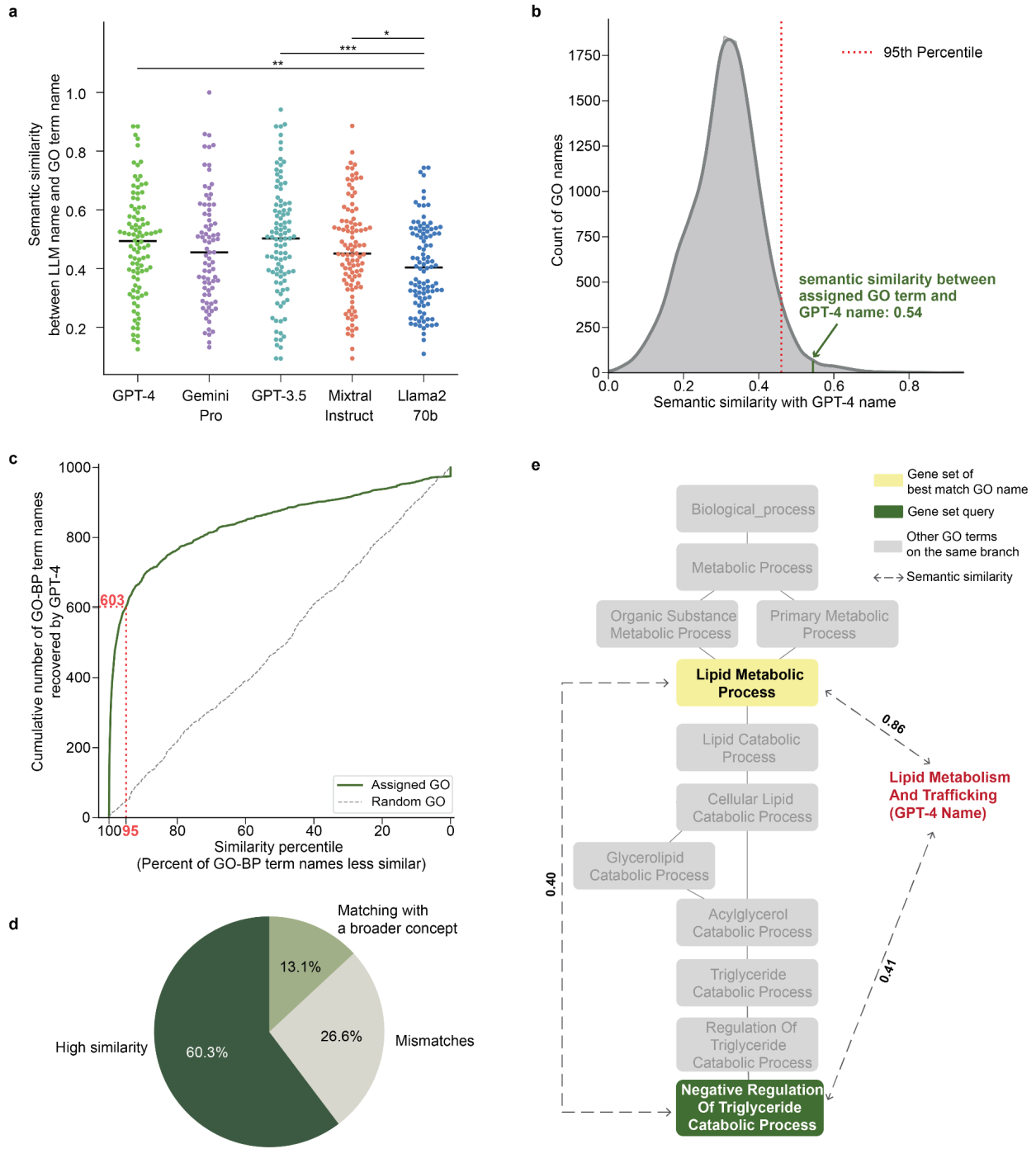

**Fig. 2: Evaluation of LLMs in recovering GO gene set names. a,** Performance of each LLM (colors) is scored by the semantic similarity between its proposed name for a gene set and the name assigned by the GO curators. Results for 100 GO terms are shown (dots; black horizontal lines show median semantic similarities). Significant difference in distributions is denoted by asterisks (*p<0.05; **p<0.01; ***p<0.001) using Mann–Whitney U test. **b**, Percentile calibration of semantic similarity between the GO and GPT-4 names for a gene set, shown for the GO term "Response to X-ray" and the corresponding GPT-4 name "DNA Damage Response and Repair". The plot shows the semantic similarity between these two names

(vertical dark green line, 0.54) versus the complete distribution of semantic similarity scores between the GPT-4 name and each name in the GO Biological Process database (GO-BP, gray). The score of the GPT-4 name is converted to a percentile, i.e. the percentage of all names in GO with lower similarity (here, 99%). Red dashed line denotes the 95th percentile threshold. **c**, Cumulative number of GO term names recovered by GPT-4 (y-axis) at a given similarity percentile (x-axis). 0 = least similar, 100 = most similar. Dark green curve: semantic similarities between GPT-4 names and assigned GO term names. Grey dashed curve: semantic similarities between GPT-4 names and random GO term names. The red dotted line marks that 603 of 1000 sampled GO names are recovered by GPT-4 at the 95th similarity percentile. **d**, Pie chart summarizing the results of the GPT-4 name / GO name similarity comparison. **e**, Hierarchical view of the GO term "Negative Regulation of Triglyceride Catabolic Process" and its ancestors. Blue box: gene set query, yellow box: gene set of best match GO name (most similar GO name to GPT-4 name), dashed lines with arrows: semantic similarities between names, red text: GPT-4 proposed name.

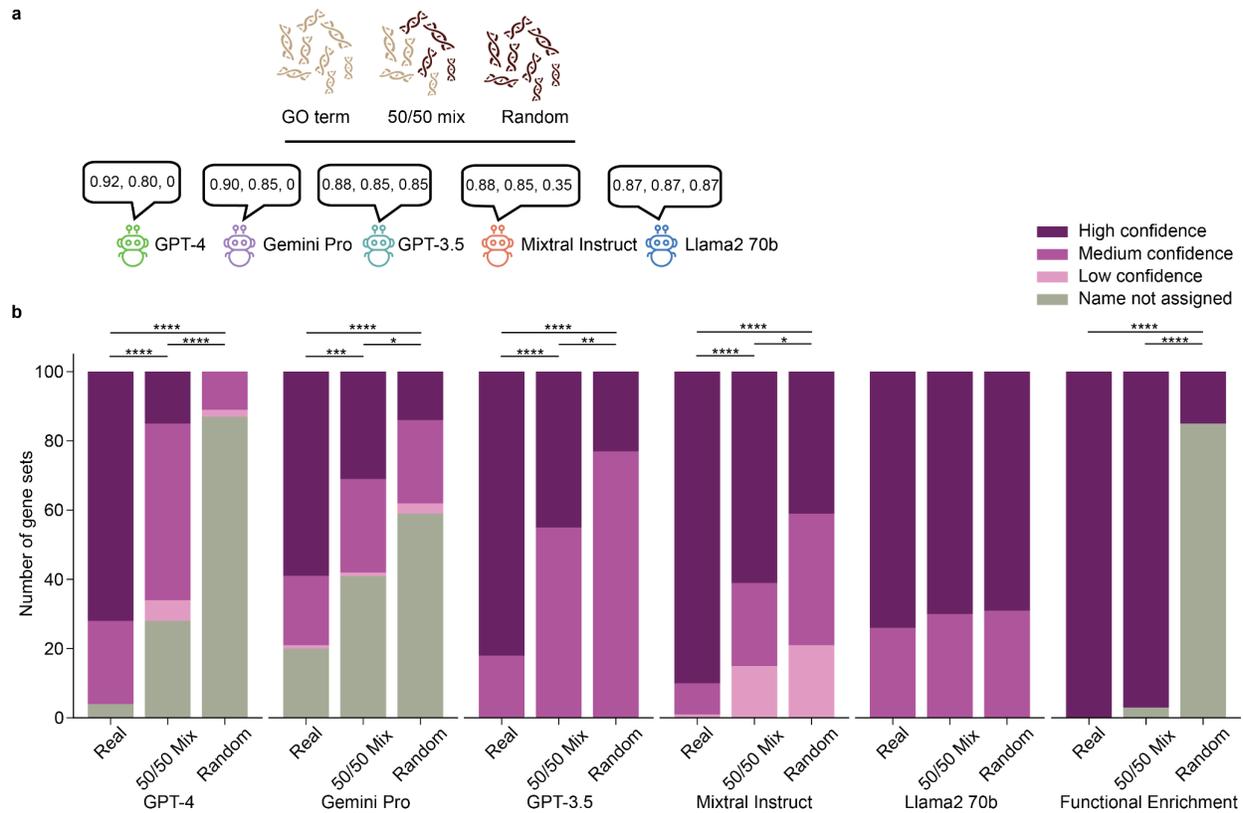

**Fig. 3: Evaluation of LLM self-confidence. a,** Investigation of model-assigned confidence scores (chat bubbles) for the ability to distinguish actual GO terms from 50/50 mix and random gene sets (light DNA strands from the same GO term, dark DNA strands randomly selected from outside the GO term). **b,** Bar graphs showing the confidence rating assigned by each model for real, contaminated, or random gene sets. Increasing shades of purple indicate low to high score bins. "High confidence" (dark purple): 0.87–1.00; "Medium confidence" (medium purple): 0.80–0.86; "Low confidence" (light purple): 0.01–0.79; and "Name not assigned" (gray): 0. For comparison to functional enrichment (rightmost group of bars), "High confidence" for a gene set is defined as p ≤ 0.05 (dark purple, Benjamini-Hochberg correction), otherwise "Name not assigned" (gray) is used. Significant difference in confidence distributions between real, 50/50 mix and random is denoted by asterisks (*p<0.05; **p<0.01; ***p<0.001, ****p<0.0001) using chi-squared test.

**Table 1: Best and worst LLM names for GO terms by semantic similarity.**

| GO Name (GO term ID) | LLM Name | Semantic Similarity | LLM |
|---|---|---|---|
| *LLM Names with Highest Similarity to GO Names* | | | |
| Synaptic vesicle exocytosis (GO:0016079) | Synaptic vesicle exocytosis | 1.00 | Gemini Pro |
| Synaptic vesicle exocytosis (GO:0016079) | Synaptic vesicle exocytosis and neurotransmitter release | 0.94 | GPT-3.5 |
| Pentose-phosphate shunt (GO:0006098) | Pentose phosphate pathway | 0.89 | GPT-3.5 |
| Glucose-6-phosphate transport (GO:0015760) | Glucose-6-phosphate metabolism and transport | 0.89 | Mixtral Instruct |
| Protein quality control for misfolded or incompletely synthesized proteins (GO:0006515) | Protein quality control and degradation | 0.88 | GPT-4 |
| *LLM Names with Lowest Similarity to GO Names* | | | |
| negative regulation of fat cell differentiation (GO:0045599) | Regulation of Wnt signaling and cellular stress response | 0.13 | GPT-4 |
| Negative regulation of CD8-positive, alpha-beta T cell differentiation (GO:0043377) | Regulation of iron homeostasis | 0.11 | Llama2 70b |
| Negative regulation of peptide secretion (GO:0002792) | Glucose homeostasis and energy metabolism | 0.09 | GPT-3.5 |
| Negative regulation of peptide secretion (GO:0002792) | Glucose homeostasis and energy metabolism | 0.09 | Mixtral Instruct |
| Negative regulation of CD8-positive, alpha-beta T cell differentiation (GO:0043377) | Regulation of ion transport and cellular homeostasis | 0.09 | GPT-3.5 |

**Table 2: Numbers of 'omics gene sets annotated by GPT-4 versus functional enrichment.**

| | | GPT-4 | | Total |
|---|---|---|---|---|
| | | Annotated † | Not Annotated | |
| **Functional Enrichment** | Annotated ‡ | 36 | 4 | 40 |
| | Not Annotated | 97 | 163 | 260 |
| | Total | 133 | 167 | **300** |

† A gene set is annotated via GPT-4 if the confidence score ≥ 0.80
‡ A gene set is annotated via enrichment analysis if BH-adjusted p ≤ 0.05 and Jaccard Index ≥ 0.10.

**Table 3: Representative analysis for protein interaction clusters (NeST: "Cluster2-123").**

| Genes (n = 13) | LYST NRDE2 REN SNRPA1 SNX6 SNX7 SNX12 SORT1 TMEM9 TMEM123 TMEM189 WASHC3 WDFY1 | |
|---|---|---|
| **Analysis Method** | **Assigned Name** | **Confidence / Enrichment FDR (JI)** |
| GPT-4 | Endosomal Sorting and Trafficking | 0.85 (Medium) |
| Gene set enrichment | Negative Regulation of Lipoprotein Lipase Activity (GO:0051005) | 0.036 (0.059) |

**GPT-4 Analysis Paragraph**

1. LYST, or lysosomal trafficking regulator, is involved in the regulation of lysosomal size and morphology. It plays a role in the sorting and trafficking of endosomal cargo destined for lysosomes, which is crucial for maintaining cellular homeostasis and proper organelle function.

> Westphal et al. "Lysosomal trafficking regulator Lyst links membrane trafficking to toll-like receptor-mediated inflammatory responses." The Journal of experimental medicine, 2017, pp. 227-244.

> Setaluri. "Sorting and targeting of melanosomal membrane proteins: signals, pathways, and mechanisms." Pigment cell research, 2000, pp. 128-34.

> Steffens et al. "Physical, Functional and Genetic Interactions between the BEACH Domain Protein SPIRRIG and LIP5 and SKD1 and Its Role in Endosomal Trafficking to the Vacuole in Arabidopsis." Frontiers in plant science, 2017, pp. 1969.

2. NRDE2 is less characterized but is thought to be involved in nuclear RNAi and may play a role in the regulation of gene expression at the post-transcriptional level, potentially influencing the stability or localization of specific mRNA transcripts.

> Wang et al. "NRDE2 negatively regulates exosome functions by inhibiting MTR4 recruitment and exosome interaction." Genes & development, 2019, pp. 536-549.

> Wan et al. "A Conserved NRDE-2/MTR-4 Complex Mediates Nuclear RNAi in Caenorhabditis elegans." Genetics, 2020, pp. 1071-1085.

> Flemr et al. "Mouse nuclear RNAi-defective 2 promotes splicing of weak 5' splice sites." RNA (New York, N.Y.), 2023, pp. 1140-1165.

3. REN, also known as renin, is primarily involved in blood pressure control and electrolyte balance. It catalyzes the conversion of angiotensinogen to angiotensin I and is a key component of the renin-angiotensin system (RAS), which is not directly related to endosomal sorting but may be affected by vesicular trafficking processes.

> Gubler et al. "Renin-angiotensin system in kidney development: renal tubular dysgenesis." Kidney international, 2010, pp. 400-6.

> Rossi et al. "Interactions between endothelin-1 and the renin-angiotensin-aldosterone system." Cardiovascular research, 1999, pp. 300-7.

> Liu et al. "AMPK: a balancer of the renin-angiotensin system." Bioscience reports, 2019.

4. SNRPA1 is a small nuclear ribonucleoprotein that is part of the spliceosome complex. It is involved in pre-mRNA splicing, a process that occurs in the nucleus and is not directly related to endosomal sorting.

    Wu et al. "Major spliceosome defects cause male infertility and are associated with nonobstructive azoospermia in humans." Proceedings of the National Academy of Sciences of the United States of America, 2016, pp. 4134-9.

    Tanikawa et al. "The spliceosome U2 snRNP factors promote genome stability through distinct mechanisms; transcription of repair factors and R-loop processing." Oncogenesis, 2016, pp. e280.

    Kim et al. "The unique spliceosome signature of human pluripotent stem cells is mediated by SNRPA1, SNRPD1, and PNN." Stem cell research, 2017, pp. 43-53.

5. SNX6, SNX7, and SNX12 are members of the sorting nexin family. These proteins are involved in endosomal sorting and trafficking, playing roles in the retromer complex, which is responsible for the retrograde transport of cargo from endosomes to the trans-Golgi network.

    Bonifacino and Hurley. "Retromer." Current opinion in cell biology, 2008, pp. 427-36.

    Harterink et al. "A SNX3-dependent retromer pathway mediates retrograde transport of the Wnt sorting receptor Wntless and is required for Wnt secretion." Nature cell biology, 2011, pp. 914-923.

    Wassmer et al. "The retromer coat complex coordinates endosomal sorting and dynein-mediated transport, with carrier recognition by the trans-Golgi network." Developmental cell, 2009, pp. 110-22.

6. SORT1, or sortilin, functions in the sorting of proteins to the lysosomes. It is involved in the transport of lysosomal enzymes and may also play a role in neurotrophic signaling and apoptosis.

    Jansen et al. "Roles for the pro-neurotrophin receptor sortilin in neuronal development, aging and brain injury." Nature neuroscience, 2007, pp. 1449-57.

    Strong et al. "Hepatic sortilin regulates both apolipoprotein B secretion and LDL catabolism." The Journal of clinical investigation, 2012, pp. 2807-16.

    Sánchez et al. "SORT1 Mutation Resulting in Sortilin Deficiency and p75(NTR) Upregulation in a Family With Essential Tremor." ASN neuro, 2015.

7. TMEM9, TMEM123, and TMEM189 are transmembrane proteins with less well-characterized functions. However, transmembrane proteins often play roles in cellular transport and signaling, potentially contributing to the trafficking system.

    Jung et al. "TMEM9 promotes intestinal tumorigenesis through vacuolar-ATPase-activated Wnt/β-catenin signalling." Nature cell biology, 2018, pp. 1421-1433.

    Dodeller et al. "The lysosomal transmembrane protein 9B regulates the activity of inflammatory signaling pathways." The Journal of biological chemistry, 2008, pp. 21487-94.

    Jung et al. "TMEM9-v-ATPase Activates Wnt/β-Catenin Signaling Via APC Lysosomal Degradation for Liver Regeneration and Tumorigenesis." Hepatology (Baltimore, Md.), 2021, pp. 776-794.

8. WASHC3 is part of the WASH complex, which is involved in endosomal sorting and the regulation of actin polymerization. It plays a role in the fission of transport intermediates from the endosomes.

9. WDFY1 is a WD repeat and FYVE domain-containing protein that is implicated in endosomal sorting and signaling pathways, potentially acting as an adaptor in the endosomal system.

    Hu et al. "WDFY1 mediates TLR3/4 signaling by recruiting TRIF." EMBO reports, 2015, pp. 447-55.

    Yeo et al. "PRDX6 Inhibits Neurogenesis through Downregulation of WDFY1-Mediated TLR4 Signal." Molecular neurobiology, 2019, pp. 3132-3144.

    Ning et al. "Hes1 attenuates type I IFN responses via VEGF-C and WDFY1." The Journal of experimental medicine, 2019, pp. 1396-1410.

> To summarize, the majority of the proteins in this set are involved in the processes of endosomal sorting and trafficking, which are essential for the proper distribution and degradation of cellular components. While not all proteins are directly linked to this process, the preponderance of sorting nexins and related proteins suggests a strong functional theme. The presence of proteins like REN and SNRPA1, which are not directly related to endosomal sorting, slightly lowers the confidence score. However, the overall function of the system appears to be centered on the endosomal-lysosomal pathway and the regulation of cargo sorting and trafficking within the cell.
>
> Xu et al. "Lysosomal physiology." Annual review of physiology, 2015, pp. 57-80.

† Orange text: unverified statement with unassertive tone

## References

1.  Zeeberg, B. R. *et al.* GoMiner: a resource for biological interpretation of genomic and proteomic data. *Genome Biol.* **4**, R28 (2003).

2.  Breitling, R., Amtmann, A. & Herzyk, P. Iterative Group Analysis (iGA): a simple tool to enhance sensitivity and facilitate interpretation of microarray experiments. *BMC Bioinformatics* **5**, 34 (2004).

3.  Beissbarth, T. & Speed, T. P. GOstat: find statistically overrepresented Gene Ontologies within a group of genes. *Bioinformatics* **20**, 1464–1465 (2004).

4.  Subramanian, A. *et al.* Gene set enrichment analysis: a knowledge-based approach for interpreting genome-wide expression profiles. *Proc. Natl. Acad. Sci. U. S. A.* **102**, 15545–15550 (2005).

5.  Al-Shahrour, F. *et al.* From genes to functional classes in the study of biological systems. *BMC Bioinformatics* **8**, 114 (2007).

6.  Backes, C. *et al.* GeneTrail—advanced gene set enrichment analysis. *Nucleic Acids Res.* **35**, W186–W192 (2007).

7.  Huang, D. W., Sherman, B. T. & Lempicki, R. A. Systematic and integrative analysis of large gene lists using DAVID bioinformatics resources. *Nat. Protoc.* **4**, 44–57 (2009).

8.  Chen, E. Y. *et al.* Enrichr: interactive and collaborative HTML5 gene list enrichment analysis tool. *BMC Bioinformatics* **14**, 128 (2013).

9.  Pomaznoy, M., Ha, B. & Peters, B. GOnet: a tool for interactive Gene Ontology analysis. *BMC Bioinformatics* **19**, 470 (2018).

10. Cerami, E. G. *et al.* Pathway Commons, a web resource for biological pathway data. *Nucleic Acids Res.* **39**, D685–90 (2011).

11. Fabregat, A. *et al.* The Reactome pathway Knowledgebase. *Nucleic Acids Res.* **44**, D481–7 (2015).

12. Pico, A. R. *et al.* WikiPathways: pathway editing for the people. *PLoS Biol.* **6**, e184 (2008).

13. Kanehisa, M., Goto, S., Sato, Y., Furumichi, M. & Tanabe, M. KEGG for integration and interpretation of large-scale molecular data sets. *Nucleic Acids Res.* **40**, D109–14 (2012).

14. Pillich, R. T. *et al.* NDEx IQuery: a multi-method network gene set analysis leveraging the Network Data Exchange. *Bioinformatics* **39**, (2023).

15. Wang, S. *et al.* Typing tumors using pathways selected by somatic evolution. *Nat. Commun.* **9**, 4159 (2018).

16. Ashburner, M. *et al.* Gene Ontology: tool for the unification of biology. *Nat. Genet.* **25**, 25–29 (2000).

17. Gene Ontology Consortium *et al.* The Gene Ontology knowledgebase in 2023. *Genetics* **224**, (2023).

18. Kanehisa, M. & Goto, S. KEGG: kyoto encyclopedia of genes and genomes. *Nucleic Acids Res.* **28**, 27–30 (2000).

19. Kanehisa, M. Toward understanding the origin and evolution of cellular organisms. *Protein Sci.* **28**, 1947–1951 (2019).

20. Kanehisa, M., Furumichi, M., Sato, Y., Kawashima, M. & Ishiguro-Watanabe, M. KEGG for taxonomy-based analysis of pathways and genomes. *Nucleic Acids Res.* **51**, D587–D592 (2023).

21. Croft, D. Reactome: A database of biological pathways. *Nature Precedings* (2010) doi:10.1038/npre.2010.5025.1.

22. Jassal, B. *et al.* The reactome pathway knowledgebase. *Nucleic Acids Res.* **48**, D498–D503 (2020).

23. Sollis, E. *et al.* The NHGRI-EBI GWAS Catalog: knowledgebase and deposition resource. *Nucleic Acids Res.* **51**, D977–D985 (2023).

24. Blake, J. A. *et al.* The Mouse Genome Database genotypes::phenotypes. *Nucleic Acids Res.* **37**, D712–9 (2009).

25. Weng, M.-P. & Liao, B.-Y. MamPhEA: a web tool for mammalian phenotype enrichment analysis. *Bioinformatics* **26**, 2212–2213 (2010).

26. Keenan, A. B. *et al.* ChEA3: transcription factor enrichment analysis by orthogonal omics integration. *Nucleic Acids Res.* **47**, W212–W224 (2019).

27. Rubin, J. D. *et al.* Transcription factor enrichment analysis (TFEA) quantifies the activity of multiple transcription factors from a single experiment. *Commun Biol* **4**, 661 (2021).

28. Franzén, O., Gan, L.-M. & Björkegren, J. L. M. PanglaoDB: a web server for exploration of mouse and human single-cell RNA sequencing data. *Database* **2019**, (2019).

29. Zhang, X. *et al.* CellMarker: a manually curated resource of cell markers in human and mouse. *Nucleic Acids Res.* **47**, D721–D728 (2019).

30. Hu, C. *et al.* CellMarker 2.0: an updated database of manually curated cell markers in human/mouse and web tools based on scRNA-seq data. *Nucleic Acids Res.* **51**, D870–D876 (2023).

31. Vaswani, A. *et al.* Attention is all you need. Preprint at https://doi.org/10.48550/arXiv.1706.03762 (2017).

32. Devlin, J., Chang, M.-W., Lee, K. & Toutanova, K. BERT: Pre-training of Deep Bidirectional Transformers for Language Understanding. in *Proceedings of the 2019 Conference of the North American Chapter of the Association for Computational Linguistics: Human Language Technologies, Volume 1 (Long and Short Papers)* 4171–4186 (Association for Computational Linguistics, 2019).

33. Brown, T. B. *et al.* Language Models are Few-Shot Learners. Preprint at https://doi.org/10.48550/arXiv.2005.14165 (2020).

34. OpenAI. GPT-4 Technical Report. Preprint at https://doi.org/10.48550/arXiv.2303.08774 (2023).

35. Touvron, H. *et al.* Llama 2: Open Foundation and Fine-Tuned Chat Models. Preprint at https://doi.org/10.48550/arXiv.2307.09288 (2023).

36. Jiang, A. Q. *et al.* Mixtral of Experts. Preprint at https://doi.org/10.48550/arXiv.2401.04088 (2024).

37. Gemini Team *et al.* Gemini: A Family of Highly Capable Multimodal Models. Preprint at https://doi.org/10.48550/arXiv.2312.11805 (2023).

38. Moghaddam, S. R. & Honey, C. J. Boosting Theory-of-Mind Performance in Large Language Models via Prompting. Preprint at https://doi.org/10.48550/arXiv.2304.11490 (2023).

39. Hebenstreit, K., Praas, R., Kiesewetter, L. P. & Samwald, M. An automatically discovered chain-of-thought prompt generalizes to novel models and datasets. Preprint at https://doi.org/10.48550/arXiv.2305.02897 (2023).

40. Caufield, J. H. *et al.* Structured prompt interrogation and recursive extraction of semantics (SPIRES):

A method for populating knowledge bases using zero-shot learning. Preprint at https://doi.org/10.48550/arXiv.2304.02711 (2023).

41. Wei, J. *et al.* Chain-of-Thought Prompting Elicits Reasoning in Large Language Models. Preprint at https://doi.org/10.48550/arXiv.2201.11903 (2022).

42. Miller, G. A. & Charles, W. G. Contextual correlates of semantic similarity. *Lang. Cogn. Process.* **6**, 1–28 (1991).

43. Xiong, M. *et al.* Can LLMs Express Their Uncertainty? An Empirical Evaluation of Confidence Elicitation in LLMs. Preprint at https://doi.org/10.48550/arXiv.2306.13063 (2023).

44. Fu, J., Ng, S.-K., Jiang, Z. & Liu, P. GPTScore: Evaluate as You Desire. Preprint at https://doi.org/10.48550/arXiv.2302.04166 (2023).

45. Duan, Q. *et al.* LINCS Canvas Browser: interactive web app to query, browse and interrogate LINCS L1000 gene expression signatures. *Nucleic Acids Res.* **42**, W449–60 (2014).

46. Barrett, T. *et al.* NCBI GEO: archive for functional genomics data sets--update. *Nucleic Acids Res.* **41**, D991–5 (2013).

47. Zheng, F. *et al.* Interpretation of cancer mutations using a multiscale map of protein systems. *Science* **374**, eabf3067 (2021).

48. Groh, B. S. *et al.* The antiobesity factor WDTC1 suppresses adipogenesis via the CRL4WDTC1 E3 ligase. *EMBO Rep.* **17**, 638–647 (2016).

49. Hu, Y.-H. *et al.* WDFY1 mediates TLR3/4 signaling by recruiting TRIF. *EMBO Rep.* **16**, 447–455 (2015).

50. Ning, F. *et al.* Hes1 attenuates type I IFN responses via VEGF-C and WDFY1. *J. Exp. Med.* **216**, 1396–1410 (2019).

51. Yeo, I. J. *et al.* PRDX6 Inhibits Neurogenesis through Downregulation of WDFY1-Mediated TLR4 Signal. *Mol. Neurobiol.* **56**, 3132–3144 (2019).

52. Schick, T. *et al.* Toolformer: Language Models Can Teach Themselves to Use Tools. Preprint at https://doi.org/10.48550/arXiv.2302.04761 (2023).

53. Yao, S. *et al.* ReAct: Synergizing reasoning and acting in language models. Preprint at https://doi.org/10.48550/arXiv.2210.03629 (2022).

54. Fezari, M. & Ali-Al-Dahoud, A. A. D. From GPT to AutoGPT: a Brief Attention in NLP Processing using DL. *researchgate.net*.

55. Shen, Y. *et al.* HuggingGPT: Solving AI Tasks with ChatGPT and its Friends in Hugging Face. https://doi.org/10.48550/arXiv.2303.17580 Preprint at (2023).

56. Nair, V., Schumacher, E., Tso, G. & Kannan, A. DERA: Enhancing Large Language Model Completions with Dialog-Enabled Resolving Agents. Preprint at https://doi.org/10.48550/arXiv.2303.17071 (2023).

57. Shinn, N. *et al.* Reflexion: Language Agents with Verbal Reinforcement Learning. Preprint at https://doi.org/10.48550/arXiv.2303.11366 (2023).

58. Li, G., Al Kader Hammoud, H. A., Itani, H., Khizbullin, D. & Ghanem, B. CAMEL: Communicative Agents for 'Mind' Exploration of Large Scale Language Model Society. Preprint at https://doi.org/10.48550/arXiv.2303.17760 (2023).

59. Keskar, N. S., McCann, B., Varshney, L. R., Xiong, C. & Socher, R. CTRL: A Conditional Transformer Language Model for Controllable Generation. Preprint at https://doi.org/10.48550/arXiv.1909.05858 (2019).

60. Holtzman, A., Buys, J., Du, L., Forbes, M. & Choi, Y. The Curious Case of Neural Text Degeneration. Preprint at https://doi.org/10.48550/arXiv.1904.09751 (2019).

61. Smith, B. *et al.* The OBO Foundry: coordinated evolution of ontologies to support biomedical data integration. *Nat. Biotechnol.* **25**, 1251–1255 (2007).

62. Tirmizi, S. H. *et al.* Mapping between the OBO and OWL ontology languages. *J. Biomed. Semantics* **2 Suppl 1**, S3 (2011).

63. Liu, F., Shareghi, E., Meng, Z., Basaldella, M. & Collier, N. Self-Alignment Pretraining for Biomedical Entity Representations. Preprint at https://doi.org/10.48550/arXiv.2010.11784 (2020).

64. Fang, Z., Liu, X. & Peltz, G. GSEApy: a comprehensive package for performing gene set enrichment analysis in Python. *Bioinformatics* **39**, (2023).

## Acknowledgments

We would like to thank all members of the Ideker lab, especially Xiaoyu Zhao and Akshat Singhal for proofreading the manuscript and providing insightful comments, Christopher Churas for helping with the organization of the GitHub repository. This work was supported by National Institutes of Health grants U24 CA269436, OT2 OD032742, 5U24HG012107, U01 MH115747, and U54 CA274502. Additional support was received from Schmidt Futures.

## Authors and Affiliations

**Department of Medicine, University of California San Diego, La Jolla, California, USA**

Mengzhou Hu, Ingoo Lee, Rudolf T. Pillich, Robin Bachelder, Trey Ideker & Dexter Pratt

**Department of Computer Science and Engineering, University of California San Diego, La Jolla, California, USA**

Sahar Alkhairy & Trey Ideker

**Department of Physics, University of California San Diego, La Jolla, California, USA**

Kevin Smith

## Corresponding Authors

Correspondence to Trey Ideker (tideker@ucsd.edu) and Dexter Pratt (depratt@ucsd.edu).

## Author Contributions

M.H., S.A., T.I., and D.P. designed the study. M.H. and S.A. developed and implemented the automated LLM-based gene set interpretation pipeline, performed the data analysis, and organized the GitHub repository. S.A. developed and assessed the semantic similarity calculation. I.L. and M.H. contributed to the development of the citation search and validation pipeline. D.P. contributed to the coding and the evaluation of the analysis. R.P. assisted in the study design, prompt engineering and the evaluation of the analysis. M.H., R.P., R.B. and D.P. conducted the scientific review of the LLM output. D.F. built the web interface for the GSAI tool, and K.S. set up the server for open-source LLMs. M.H., S.A., T.I., and D.P. wrote the manuscript with input from all authors. All authors approved the final version of this manuscript.

## Author Declarations

TI is a co-founder, member of the advisory board, and has an equity interest in Data4Cure and Serinus Biosciences. TI is a consultant for and has an equity interest in Ideaya Biosciences. The terms of these arrangements have been reviewed and approved by the University of California San Diego in accordance with its conflict-of-interest policies.

## Data availability

Publicly available gene sets were used in this study. Gene Ontology (2023-11-15 release) is available at http://release.geneontology.org/2023-11-15/ontology/index.html. The selected NeST gene set is available to download from https://github.com/idekerlab/llm_evaluation_for_gene_set_interpretation/blob/main/data/Omics_data/NeST__IAS_clixo_hidef_Nov17.edges. The L1000 data was downloaded from https://maayanlab.cloud/Harmonizome/dataset/LINCS+L1000+CMAP+Signatures+of+Differentially+Expressed+Genes+for+Small+Molecules. The viral infection data was downloaded from https://maayanlab.cloud/Harmonizome/dataset/GEO+Signatures+of+Differentially+Expressed+Genes+for+Viral+Infections.

## Code availability

The code for running the LLM gene set analysis pipeline and the evaluation tasks is available at https://github.com/idekerlab/llm_evaluation_for_gene_set_interpretation with the MIT License.

# Extended Data Figures and Tables

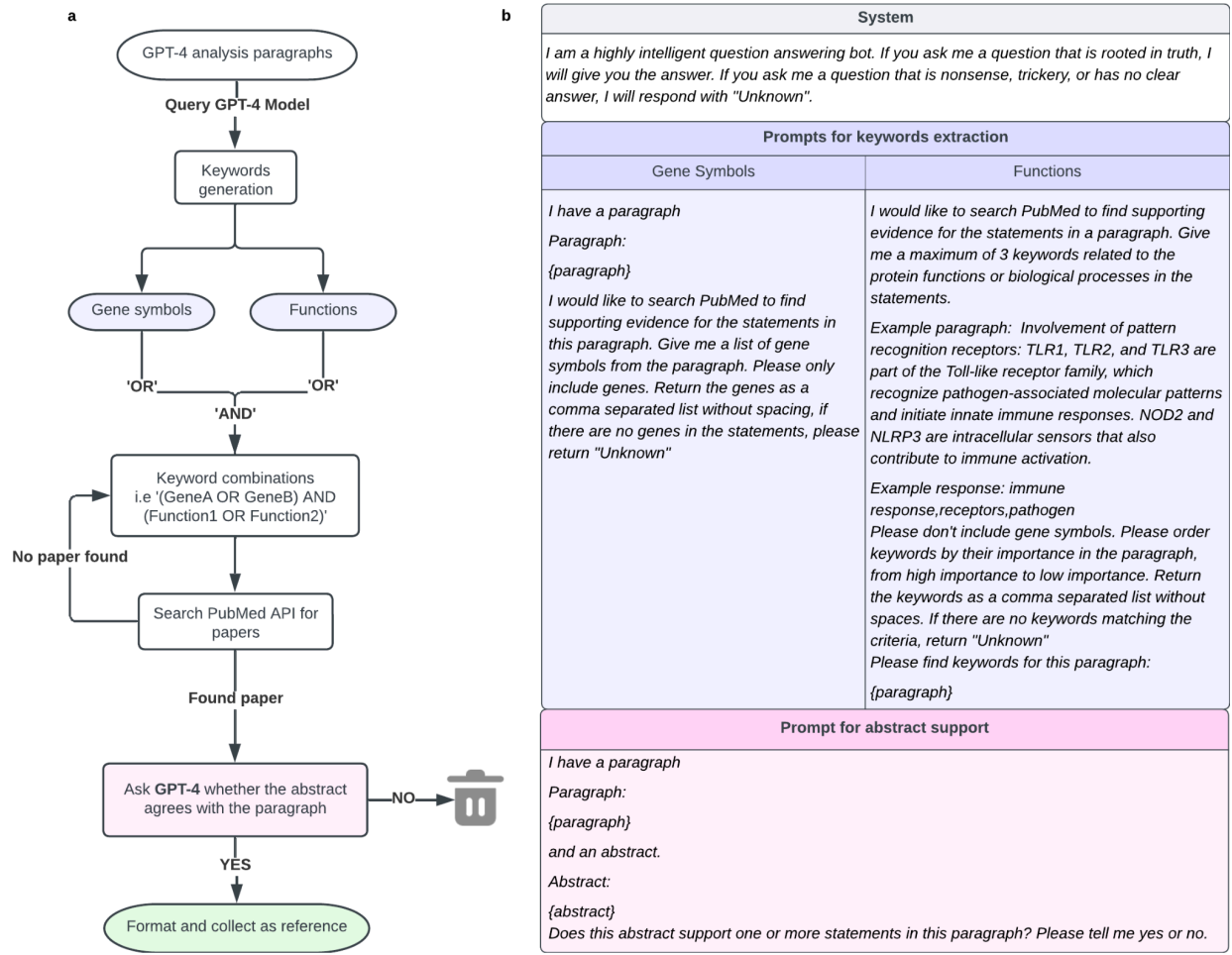

**Extended Data Fig. 1: Schematic of the citation module. a**, GPT-4 is asked to provide gene symbol keywords and functional keywords separately. Multiple gene keywords and functions are combined and used to search PubMed for relevant paper titles and abstracts in the scientific literature. GPT-4 is queried to evaluate each abstract, saving supporting references. **b,** Prompts used to query the GPT-4 model.

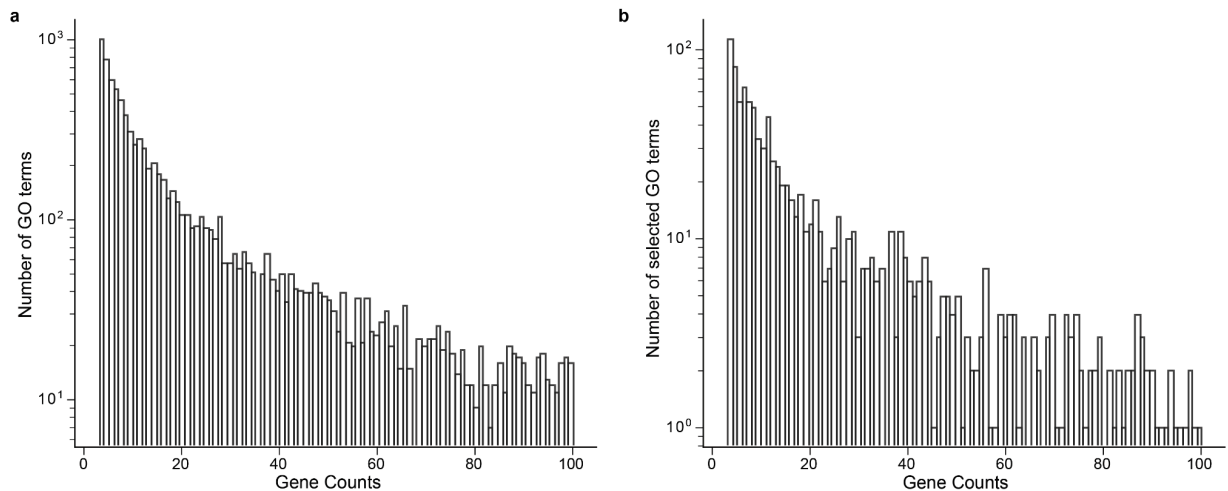

**Extended Data Fig. 2: Distribution of GO term gene sizes**. **a,** Distribution of term size (number of genes) for terms in the Biological Process branch (GO-BP). Terms with 3-100 genes shown (n = 8,910). **b,** Distribution of term size for the 1000 GO terms used in Task 1.

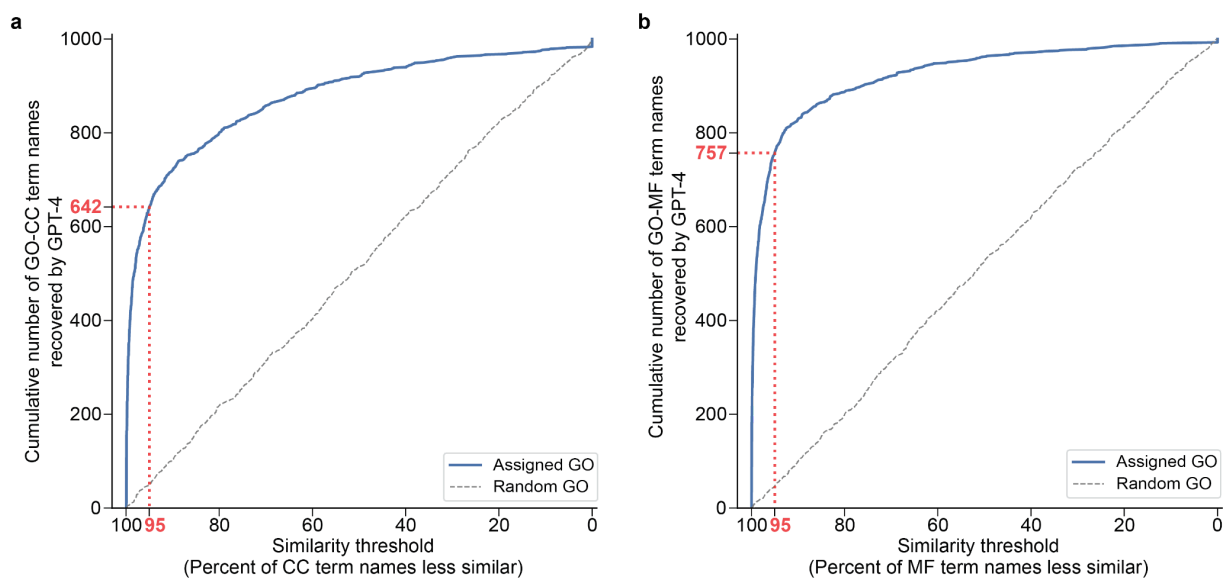

**Extended Data Fig. 3: Evaluation of GPT-4 in recovery of GO-CC and GO-MF names. a**, Cumulative number of GO-CC term names recovered by GPT-4 (y-axis) at a given similarity percentile (x-axis). 0 = least similar, 100 = most similar. Blue curve: semantic similarities between GPT-4 names and assigned GO-CC term names. Grey dashed curve: semantic similarities between GPT-4 names and random GO-CC term names. The red dotted line marks that 642 of the 1000 sampled GO-CC names are recovered by GPT-4 at a similarity percentile of 95%. **b**, As for panel a, but for GO-MF terms rather than GO-CC. The red dotted line marks that 757 of the 1000 sampled GO-MF names are recovered by GPT-4 at a similarity percentile of 95%.

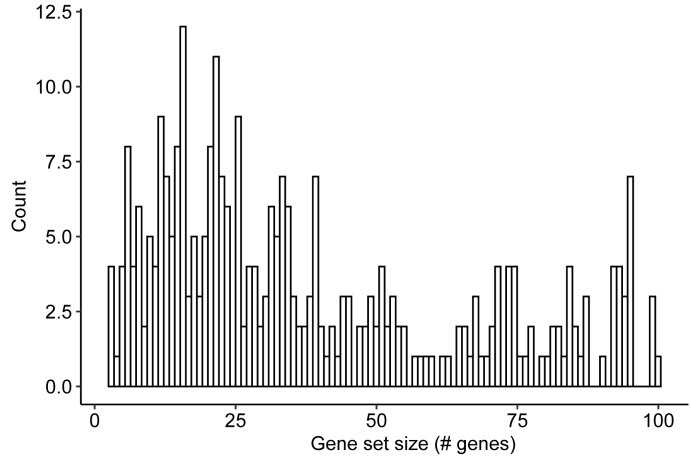

**Extended Data Fig. 4: Distribution of 'omics gene set sizes.** Distribution shown for all 'omics gene sets considered in this study (n = 300).

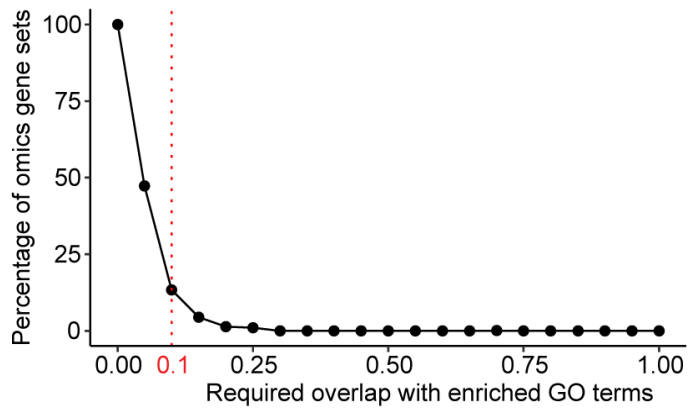

**Extended Data Fig. 5: Evaluation of required overlap.** The percentage of omics gene sets (y-axis) matched to GO terms with the required overlap (Jaccard Index, x-axis). The vertical red dashed line marks a threshold Jaccard Index = 0.1.

**Extended Data Table 1: Engineered prompt for gene set analysis.**

You are an efficient and insightful assistant to a molecular biologist

Write a critical analysis of the biological processes performed by this system of interacting proteins.
Base your analysis on prior knowledge available in your training data.
After completing your analysis, propose a brief and detailed name for the most prominent biological process performed by the system.

After completing your analysis, please also assign a confidence score to the process name you selected. This score should follow the name in parentheses and range from 0.00 to 1.00. A score of 0.00 indicates the lowest confidence, while 1.00 reflects the highest confidence. This score helps gauge how accurately the chosen name represents the functions and activities within the system of interacting proteins. When determining your score, consider the proportion of genes in the protein system that participate in the identified biological process. For instance, if you select "Ribosome biogenesis" as the process name but only a few genes in the system contribute to this process, the score should be lower compared to a scenario where a majority of the genes are involved in "Ribosome biogenesis".

Put your chosen name at the top of the analysis as 'Process: <name>'.

Be concise, do not use unnecessary words.
Be factual, do not editorialize.
Be specific, avoid overly general statements such as 'the proteins are involved in various cellular processes'.
Avoid listing facts about individual proteins. Instead, try to group proteins with similar functions and discuss their interplay, synergistic or antagonistic effects and functional integration within the system.
Also avoid choosing generic process names such as 'Cellular Signaling and Regulation'.
If you cannot identify a prominent biological process for the proteins in the system, I want you to communicate this in you analysis and name the process: "System of unrelated proteins". Provide a score of 0.00 for a "System of unrelated proteins".

To help you in your work, I am providing an example system of interacting proteins and the corresponding example analysis output.

The example system of interacting proteins is:
PDX1, SLC2A2, NKX6-1, GLP1, GCG.

The example analysis output is:

Process: Pancreatic development and glucose homeostasis (0.96)

1. PDX1 is a homeodomain transcription factor involved in the specification of the early pancreatic epithelium and its subsequent differentiation.
It activates the transcription of several genes including insulin, somatostatin, glucokinase and glucose transporter type 2.
It is essential for maintenance of the normal hormone-producing phenotype in the pancreatic beta-cell.
In pancreatic acinar cells, it forms a complex with PBX1b and MEIS2b and mediates the activation of the ELA1 enhancer.

2. NKX6-1 is also a transcription factor involved in the development of pancreatic beta-cells during the secondary transition.
Together with NKX2-2 and IRX3, controls the generation of motor neurons in the neural tube and belongs to the neural progenitor factors induced by Sonic Hedgehog (SHH) signals.

3.GCG and GLP1, respectively glucagon and glucagon-like peptide 1, are involved in glucose metabolism and homeostasis.
GCG raises blood glucose levels by promoting gluconeogenesis and is the counter regulatory hormone of Insulin.
GLP1 is a potent stimulator of Glucose-Induced Insulin Secretion (GSIS). Plays roles in gastric motility and suppresses blood glucagon levels.
Promotes growth of the intestinal epithelium and pancreatic islet mass both by islet neogenesis and islet cell proliferation.

4. SLC2A2, also known as GLUT2, is a facilitative hexose transporter. In hepatocytes, it mediates bi-directional transport of glucose across the plasma membranes,
while in the pancreatic beta-cell, it is the main transporter responsible for glucose uptake and part of the cell's glucose-sensing mechanism.
It is involved in glucose transport in the small intestine and kidney too.

To summarize, the genes in this set are involved in the specification, differentiation, growth and functionality of the pancreas, with a particular emphasis on the pancreatic beta-cell. Particularly, the architecture of the pancreatic islet ensures proper glucose sensing and homeostasis via a number of different hormones and receptors that can elicit both synergistic and antagonistic effects in the pancreas itself and other peripheral tissues.

Here are the interacting proteins:

Proteins: {protein list}

† text color matches with **Fig. 1a** prompt color

**Extended Data Table 2: Overview of five language models.**

| Models | Version release | Params | Context length (tokens) | Company | Estimated time usage (second/gene set) | Estimated cost ($/gene set) |
|---|---|---|---|---|---|---|
| GPT-4 Turbo | Nov 2023 | ~1.7T | 128k | OpenAI | 36.5 ‡ | $4.8 \times 10^{-2}$ |
| Gemini Pro | Dec 2023 | Unspecified | 32k | Google | 7.9 | 0.0 |
| GPT-3.5 Turbo | Nov 2023 | ~175B | 16k | OpenAI | 9.6 | $2.8 \times 10^{-3}$ |
| Mixtral Instruct | Dec 2023 | 13B (active), 47B (total) | 32k | MistralAI | 46.4 | 0.0 † |
| Llama2 | July 2023 | 70B | 4k | Meta | 61.8 | 0.0 † |

† Does not consider the cost to host an open-source model.
‡ GPT-4 compute time was significantly shorter (1.1s) when asking for a gene set name but not further analysis.

# Supplementary Information

**Supplementary Table 1**

Complete analysis of GO terms, 50/50 mix and random for all models (related to Task1: Fig. 2 and 3)

**Supplementary Table 2**

Complete GPT-4 analysis of GO terms (related to Task 1: Fig. 3 and Table1).

**Supplementary Table 3**

Complete GPT-4 analysis of omics gene sets (related to Task 2: Table 2).

**Supplementary Table 4**

Reviewer fact-checking of GPT-4 analysis text and citation relevance (related to Taske 2: Table 3).